%% file: main.tex
\documentclass[sigconf]{acmart}
\AtBeginDocument{%
  \providecommand\BibTeX{{%
    \normalfont B\kern-0.5em{\scshape i\kern-0.25em b}\kern-0.8em\TeX}}}

\setcopyright{acmcopyright}
\copyrightyear{2023}
\acmYear{2023}
\acmDOI{XXXXXXX.XXXXXXX}

\acmConference[ICSE 2024]{46th International Conference on Software Engineering}{April 2024}{Lisbon, Portugal}

\usepackage{enumitem}
\usepackage{multirow}
\usepackage{booktabs}
\usepackage{xcolor}
\usepackage{colortbl}
\usepackage{geometry}
\usepackage{lipsum}
\usepackage{tikz}
\usetikzlibrary{shapes.multipart}
\usepackage{titlesec}
\usepackage{tabularx}
\usepackage{graphics}

\definecolor{aliceblue}{rgb}{0.94, 0.97, 1.0}

\title{Exploring ChatGPT for Toxicity Detection in GitHub}

\author{Shyamal Mishra}
\affiliation{%
  \institution{Drexel university}
  \streetaddress{Philadelphia}
  \city{Philadelphia}
  \state{Pennsylvania}
  \country{USA}}
\email{sm4825@drexel.edu}

\author{Preetha Chatterjee}
\affiliation{%
  \institution{Drexel university}
  \streetaddress{Philadelphia}
  \city{Philadelphia}
  \state{Pennsylvania}
  \country{USA}}
\email{preetha.chatterjee@drexel.edu}

\begin{document}

\begin{abstract}
Fostering a collaborative and inclusive environment is crucial for the sustained progress of open source development. However, the prevalence of negative discourse, often manifested as toxic comments, poses significant challenges to developer well-being and productivity. 
To identify such negativity in project communications, especially within large projects, automated toxicity detection models are necessary. 
To train these models effectively, we need large software engineering-specific toxicity datasets. However, such datasets are limited in availability and often exhibit imbalance (e.g., only 6 in 1000 GitHub issues are toxic)~\cite{Raman2020}, posing challenges for training effective toxicity detection models. To address this problem, we explore a zero-shot LLM (ChatGPT) that is pre-trained on massive datasets but without being fine-tuned specifically for the task of detecting toxicity in software-related text. Our preliminary evaluation indicates that ChatGPT shows promise in detecting toxicity in GitHub, and warrants further investigation. We experimented with various prompts, including those designed for justifying model outputs, thereby enhancing model interpretability and paving the way for potential integration of ChatGPT-enabled toxicity detection into developer communication channels.
%Lastly, we integrate ChatGPT with the Slack chat platform, creating a bot that holds the potential to proactively mitigate toxicity in developer communication channels. %to show potential integration of ChatGPT-enabled toxicity detection 

%This paper delves into the identification and understanding of such toxicity within the open source developer community. Through rigorous examination, we elucidate the rationale behind these adverse interactions. Furthermore, we propose alternative strategies to effectively mitigate the impact of harmful commentary, thereby promoting a healthier and more constructive collaborative landscape. Our findings provide valuable insights for both practitioners and stakeholders seeking to uphold the principles of open source development while curtailing the detriments posed by toxic behavior.
\end{abstract}

\maketitle

\input{intro}
%\input{soa}
\input{methodology}
\input{results}

\input{conclusion}

\bibliographystyle{IEEEtran}
\bibliography{icse_merged_bib}

\end{document}

%% file: intro.tex
\section{Introduction}
The open source software (OSS) development community has emerged as a powerful engine driving innovation and collaboration across various domains. The fundamental principle of collective problem solving has led to the creation of exceptional software projects. However, amidst this collaborative landscape, the prevalence of negative discourse, often manifested as toxic comments, %poses a significant challenge to the harmony and productivity of the community 
causes substantial harm on developers, diminishing their well-being, motivation, job satisfaction, and productivity~\cite{Raman2020, blog_exp1, blog_exp5, Miller2022, ferreiraSTFU2021, ferreiraHeat2022, gunawardena_destructive_2022, gachechiladze_anger_2017, Cheriyan21, Ehsani2023_FSE}. In a 2017 GitHub survey, it was found that 50\% of OSS developers experienced negative interactions. Among those, 21\% mentioned that such behavior led them to stop contributing~\cite{oss_survey}. 

Mitigating negativity is essential to cultivate healthier and more productive software environments and retain talent~\cite{egelmanPredicting2020, qiuDetecting2022, ortu2015bullies, Destefanis2016SoftwareDD}. Collaboration platforms such as GitHub has implemented preliminary measures to tackle negative behavior, such as issue locking~\cite{locking}, but such manual inspection remains labor-intensive due to the sheer volume of daily content. More recently, machine learning-based techniques for detecting toxicity in software-related text have started emerging~\cite{Raman2020, ferreiraIncivility2022, sarkerAutomated2022, egelmanPredicting2020}. However, these models often show a high false positive rate and limited generalizability across different communication types (e.g., issue comments , code reviews)~\cite{sarkerBenchmark2020}. 
Creating effective toxicity detection models is challenging due to limited open-source toxicity datasets~\cite{Miller2022, ferreiraIncivility2022, imran2022data}. Additionally, these datasets are often imbalanced, with just 6 out of 1000 issues being toxic~\cite{Raman2020}, making training effective models a hurdle.
Therefore, there is a significant gap in the development and adoption of effective automated techniques for identifying toxic conversations in software engineering platforms.

%While advanced techniques for detecting harassment in blogs and social media are available~\cite{gunasekara-nejadgholi-2018-review, bhat-etal-2021-say-yes}, studies have shown that directly applying these tools to software-related text is ineffective~\cite{Raman2020, sarkerBenchmark2020} due to the presence of technical terms and nuanced negative behavior (e.g., prevalence of sarcasm, arrogance on GitHub). 

Large Language Models (LLMs) have recently gained prominence as a powerful category of deep learning technique, demonstrating their effectiveness in related tasks such as emotion and sentiment analysis~\cite{batra2021bert, eeshita-sentiment, kamath2022emoroberta, liu2019dens, kabir2023answers, hou2023large, zhang2023sentiment, fan2023large, 10132255, Huang_2023, li2023hot, he2023prompt, ziems2023large}. 
Some of the most substantial and influential LLMs, such as ChatGPT~\cite{chatgpt}, are readily available that can be deployed as a ``zero-shot" model, without requiring specific fine-tuning for a particular task. Given that creating an extensive training dataset for  toxicity detection in software engineering communication is costly and resource-intensive, utilizing a zero-shot approach offers an attractive alternative. 
In this paper, we explore ChatGPT~\cite{chatgpt} in detecting toxicity in software-related text. We use a benchmark toxicity dataset of 1597 GitHub issue comments~\cite{Raman2020} to evaluate the model, and conduct a qualitative analysis to understand the common errors. Specifically, we investigate the following research questions:
\vspace{-0.4cm}
\begin{itemize}[leftmargin=*]
    \item RQ1. How effective is OpenAI's ChatGPT %compared to existing state-of-the-art toxicity detection tools (e.g., Perspective API, Raman et al., Sarkar et al.) 
    in detecting toxic text on GitHub?
    \item RQ2. What types of toxic comments are misclassified or difficult to detect using ChatGPT?
\end{itemize}
\vspace{-0.4cm}
Our preliminary findings suggest that ChatGPT shows promise in detecting toxicity
in GitHub, achieving a precision of 0.49 and recall of 0.94. This result is comparable to the state-of-the-art domain-specific toxicity detectors (e.g., Raman et al.~\cite{Raman2020} showed a precision of 0.91 and a recall of 0.42 on the same dataset). 
We explore different prompts, including those intended for explaining model results, thus improving model interpretability, and thus opening up possibilities for integrating ChatGPT-supported toxicity detection into developer communication platforms. We also show an example implementation of possible proactive moderation of toxic content by integrating ChatGPT into Slack, a popular developer chat platform.

%To demonstrate the effectiveness of ChatGPT in SE applications, we conduct a short case study to assess if it can be embedded in developer communication platforms to proactively mitigate toxic conversations. Specifically, we integrate ChatGPT as a bot in developer chat platforms (i.e., Slack). First, it tags toxic content. Second, it suggests alternate suggestions. Case Study: Can we use Open AI to rephrase toxic content and come up with alternate suggestions as a bot?

%In the pursuit of a comprehensive analysis, we delve into the rationale behind toxic comments, aiming to uncover the motivations and triggers that drive people to engage in harmful discourse. Furthermore, we endeavor to present alternative strategies and approaches that can effectively mitigate the impact of toxic comments while fostering a more respectful and constructive environment.

%% file: methodology.tex
\section{Methodology}

\subsection{Dataset}
We analyze a benchmark dataset of 1597 GitHub issue comments (102 toxic, 1495 non-toxic) manually curated by Raman et al. \cite{Raman2020}. First, they used the GitHub API to identify issue threads that had been locked as ``too heated" among 118k GitHub projects. Of the 294k locked issues, 654 were explicitly marked as ``too heated". Several of these locked discussions contained toxic behavior. Next, they manually reviewed these discussions, labeling comments as toxic or non-toxic.

\subsection{ OpenAI Chat Completion API} 
We employ OpenAI's GPT-3.5 Turbo through the Chat Completion API to investigate toxicity in software developer community conversations. The Chat Completion API allows us to generate responses to a fixed prompt, simulating a conversation with the model. We use this API to create a conversational setting where we can evaluate the content for potential toxicity.

\subsection{Prompt Design} 
Effective prompt design is a critical aspect of utilizing OpenAI APIs to achieve desired outcomes in natural language processing tasks. The choice of prompt significantly influences the quality, relevance, and accuracy of the model's responses. Our approach involves designing specific prompts that guide the model's responses in evaluating the level of toxicity in the provided conversation. The prompts were carefully designed by using the existing literature on this topic~\cite{stowe2022impli, zhou2021pie, haagsma2020magpie, white2023prompt}.

We conducted a series of experiments employing various prompt structures and formulations. The goal of these experiments was to identify the most effective prompt design that aligns with our research objective of detecting toxicity in software-related text. We tested 15 different prompts, each with unique characteristics, instructions, and response options.

\subsubsection{Prompt Variations.}

We explored few strategies to design the prompt to detect toxicity. Our prompt variations are guided by Google's Conversation AI research in toxicity detection~\cite{CAT}. Some of the prompt characteristics we investigated include:

\begin{itemize}[leftmargin=*]
    \item The category of toxicity labels; i.e., binary (yes/no),  multiclass (e.g., "Very Toxic," "Toxic," "Slightly Toxic")
    %\item The inclusion of explicit toxicity descriptors (e.g., "Very Toxic," "Toxic," "Slightly Toxic").
    \item The framing of the task; i.e., question-based (e.g., ``Does the following comment exhibit toxicity?"), statement-based (e.g., ``Assess the following statement for any indication of toxicity.")
    %\item The length and details of the task
\end{itemize}

%\subsubsection{Observations and Findings}

Through these experiments, we made the following observations:

\begin{itemize}[leftmargin=*]
    \item Clarity of task instruction: Prompts that provided clear and specific instructions led to more consistent and accurate assessments of toxicity. Ambiguity in prompts often resulted in less reliable outcomes.
    \item Length of task instruction: While detailed prompts were helpful in conveying the desired task, overly lengthy/complex instructions sometimes led to model confusion, and verbosity in the outputs. Striking the right balance between clarity and length was essential for better performance.
    \item Classification Category: The use of multi-level toxicity categories, such as "Very Toxic," "Toxic," and "Slightly Toxic," allowed for better understanding of  subtle or nuanced forms of toxicity (e.g., passive aggressiveness).
    
\end{itemize}

\subsubsection{Selected Prompts.}
Based on the observations of our experiments, we selected three prompts that demonstrated effectiveness in detecting toxicity (in terms of our evaluation metrics). %These prompts were employed in our research to evaluate the model's performance in detecting toxicity in software developer community discussions. 
Due to the lack of space, we present only the 3 selected prompts as follows:

\begin{table}[htbp]
    \setlength{\tabcolsep}{2pt}
    \renewcommand{\arraystretch}{1.5}
    \centering
    \caption{Prompt Descriptions for Toxicity Detection}
    \vspace{-0.2cm}
    \begin{tabular}{|p{0.15\linewidth}|p{0.75\linewidth}|}
        \hline
        %\textbf{Prompt No} & \textbf{ Prompt Descriptions} \\
        %\hline
        Prompt 1 & Given the following software developer conversation, do you think this contains any toxicity? Please provide your answer in \textit{yes/no}. \\ 
        \hline
        Prompt 2 & Examine whether the software engineering community discussion includes toxicity. Please provide your answer in \textit{Very Toxic/Toxic/Slightly Toxic/Not Toxic}. \\
        \hline
        Prompt 3 & Examine whether the open-source developer community discussion includes toxicity. 
        \newline \quad \textit{Very Toxic} - a very hateful, aggressive, disrespectful comment or otherwise very likely to make a user leave a discussion or give up on sharing their perspective. 
        \newline \quad \textit{Toxic} - a rude, disrespectful, unreasonable comment or otherwise somewhat likely to make a user leave a discussion or give up on sharing their perspective. 
        \newline \quad \textit{Slightly Toxic} or hard to say. 
        \newline \quad \textit{Not toxic.} \\
        \hline
    \end{tabular}
\end{table}
\vspace{-0.2cm}

\subsection{Model Parameters} LLM generated outputs can be non-deterministic~\cite{ouyang2023llm}. We conducted extensive experimentation with different temperature settings to observe their impact on toxicity detection. Temperature influences how this model generates text~\cite{openai_temp}. When set to a lower value, such as 0.2, the generated text becomes more deterministic, honing in on specific patterns and producing relatively conservative, "safe" responses. In contrast, increasing the temperature to higher values, such as 1.0 or above, introduces an element of randomness, bestowing the text with greater diversity and creativity. We used three temperature values; 0.2, 0.7, and 1.2 for our experiments. For the rest of the parameters, we used the default settings.

\subsection{Evaluation Metrics} We evaluate the model's performance using a set of metrics that are frequently used to evaluate classification tasks: {\em F1-score}, {\em Precision}, and {\em Recall}. These metrics provide insights into the model's ability to identify true positives (TP), false positives (FP), false negatives (FN), and true negatives (TN).
Precision is the ratio of TP instances to the total predicted positive instances (i.e., $\text{\em Precision} =\frac{TP}{TP +FP}$), and Recall is the ratio of TP instances to all instances in the positive class (i.e., $\text{\em Recall} =\frac{TP}{TP +FN}$). {\em F1-score} is the harmonic mean of {\em Precision} and {\em Recall} (i.e., $\text{\em F1-score} = 2 * \frac{Precision * Recall}{Precision + Recall}$).

For calculating the evaluation metrics, we need to convert the toxicity labels of the model outputs to binary (toxic/non-toxic). Prompt 1 outputs are already in this format. For Prompts 2 and 3, we consider \textit{Very Toxic/Toxic/Slightly Toxic} as `toxic', and not toxic as `non-toxic'.

%% file: results.tex
\definecolor{graybackground}{RGB}{220, 220, 220}

\section{Preliminary Observations} In this section, we discuss our observations in using ChatGPT to detect toxic instances of GitHub issue comments.

\subsection{Evaluating Effectiveness}

\colorbox{graybackground}{%
  \parbox{\dimexpr\linewidth-2\fboxsep}{%
    \textcolor{black}{%
      \textit{RQ1. How effective is OpenAI’s ChatGPT in detecting toxic text on GitHub?}%
    }%
  }%
}

\noindent
Table \ref{evaluation} shows the results of the three selected prompts with varying temperature values (0.2, 0.7, 1.2) in detecting toxicity in the Raman et al. GitHub issue comments dataset. 

We observe that the choice of prompts has a notable impact on the performance of the model in toxicity detection. Among all the three prompts, Prompt 1 produced most effective results, achieving the highest F-score of 0.64 with temperature=0.2, 0.55 with temperature=0.7, and 0.54 with temperature=1.2. There could be several reasons for this outcome such as: 
(a) Simplicity: Prompt 1's straightforward question simplifies the task for the model. It only requires determining whether toxicity is present or not (binary classification);
%(b) Binary Classification: Binary classification often leads to a better balance between precision and recall, making it easier for the model to achieve a high F-measure.
(b) Lack of Ambiguity: Prompt 1 avoids fine-grained distinctions between toxicity levels, reducing ambiguity in decision.

\begin{table}[htbp]
    \centering
    \caption{Results for Toxicity Detection}
    \vspace{-0.2cm}
    \label{evaluation}
    \begin{tabular}{|c|c|c|c|c|}
        \hline
        Temperature & Prompt & Precision & Recall & F-measure \\
        \hline
        \multirow{0.2}{*}{0.2} & Prompt 1 & 0.49 & 0.94 & \textbf{0.64} \\
        \cline{2-5}
        & Prompt 2 & 0.36 & 0.88 & 0.51 \\
        \cline{2-5}
        & Prompt 3 & 0.33 & 0.78 & 0.48 \\
        \hline
        \multirow{0.7}{*}{0.7} & Prompt 1 & 0.43 & 0.49 & \textbf{0.55} \\
        \cline{2-5}
        & Prompt 2 & 0.35 & 0.89 & 0.50 \\
        \cline{2-5}
        & Prompt 3 & 0.39 & 0.71 & 0.51 \\
        \hline
        \multirow{1.2}{*}{1.2} & Prompt 1 & 0.41 & 0.80 & \textbf{0.54} \\
        \cline{2-5}
        & Prompt 2 & 0.59 & 0.42 & 0.49 \\
        \cline{2-5}
        & Prompt 3 & 0.29 & 0.86 & 0.43 \\
        \hline
    \end{tabular}
\end{table}

We also notice that the choice of temperature can significantly influence the trade-off between precision and recall in toxicity detection. For our dataset, a temperature setting of 0.2 (lower temperature values produce deterministic outputs) provided the best results. More experiments are required to explore ChatGPT's model determinism. 
%This choice was motivated by the fact that lower temperature settings tend to yield outcomes that are consistently uniform, whereas higher settings foster greater diversity and creativity in the generated content, as documented in \cite{openai_temp}.

\subsection{Identifying Challenges in Toxic Comment Detection}

\colorbox{graybackground}{%
  \parbox{\dimexpr\linewidth-2\fboxsep}{%
    \textcolor{black}{%
      \textit{RQ2. What types of toxic comments are misclassified or difficult to detect using ChatGPT?}%
    }%
  }%
}

\noindent
In order to gain insight into the models’ mistakes, we conduct an error analysis on the FPs and FNs for the overall best performing configuration (Temperature=0.2, Prompt=1). \textit{False Positives (FP)} occur when the model incorrectly identifies non-toxic comments as toxic; %In other words, it flags content as toxic when it isn't.
\textit{False Negatives (FN)} occur when the model fails to identify toxic comments. %It fails to flag content as toxic when it should.

%Due to the remarkably high consensus among the OpenAI API on instances of false positives (FP), our error analysis starts with analyzing these 106 FP instances.\\

\subsubsection{False Positives (FP):} We observed a total of 100 FP instances, reflecting a low precision of 0.49.  Our manual examination reveals that the FP errors can be classified into a few primary categories, which are elaborated below:

%"Getting defensive? You know what you did." may carry a negative tone but lack explicit offensive language 

\textbf{Absence of Explicit Offensive Language:} Instances that do not contain explicit offensive words or hate speech, typical indicators of toxicity, are misinterpreted. One such example in our dataset is, "\textit{Such a plugin already exists. And you chose to use a bleeding-edge build with it removed. You have nobody to blame but yourself.}". %and "@XX I actually unlocked this issue again to give you a chance to say sorry..."

\textbf{Sarcasm and Irony:} OpenAI API often encountered challenges in accurately detecting figurative language, such as sarcasm and irony. For example, statements such as, ``\textit{Really, thanks to you, I got ruined the world of survival and now have to do the cleaning map}", %and "So basically you have an extremely easy to exploit security vulnerability and decided to only fix this in the paid Enterprise Edition?" 
containing sarcasm was misinterpreted. 

%\textbf{Complex Linguistic Structures:} The presence of complex language structures or intricate arguments in these examples can obscure toxicity from automated detection systems. For instance, "Maybe you and I have different definitions of 'seriously.'..." and "Good job, then what do you want from us? Stop mentioning me."

\textbf{Context-Dependent Toxicity:} Toxicity often depends on the context, and statements that seem non-toxic in isolation may be perceived as toxic when considered within a broader  context. Some examples are, ``I didn't file a bug for Fedora, people are aware that Fedora is an unstable system and it is a test bed for RHEL." and ``I hava a 512m vps, I want to build some website on it..."

\subsubsection{False Negatives (FN):} We observed only 6 FN instances, reflecting a high recall of 0.94.  Our manual examination reveals that most of the FN errors can be classified into one broad category as follows:

\textbf{Nuanced Toxicity:} ChatGPT faces challenges in detecting nuances forms of toxicity, such as arrogance and passive-aggressiveness. For example, a comment like, ``As I said above... Issue 87 was a harmless use-after-free on shutdown, and nothing to do with this. I've said my piece, and I can see it was a mistake trying to engage with you, so I'm locking this thread," was misclassified, despite exhibiting arrogance.

\subsubsection{Common errors in FP and FN:} Across FPs and FNs, we observed some common error patterns. 

\textbf{Lengthy Phrasing: }One factor influencing both FP and FN instances is the tokenization of text by OpenAI language models. Language models read and write text in chunks called tokens, where a token can range from a single character to an entire word. For example, the sentence "ChatGPT is great!" is encoded into six tokens: ["Chat", "G", "PT", " is", " great", "!"] \cite{openai_tokens}. This tokenization process can occasionally lead to misunderstandings, particularly when dealing with lengthy phrases or complex sentences.

\textbf{Non-Responsive Answers:} Another challenge arises when the model fails to provide answers that align with our query. For instance like, "@friend Done. Cached for , as spotted in Doctrine. Is there any relevant difference?" the model responded with, "I'm sorry, but I cannot provide a yes or no answer to this question as it requires subjective analysis of the software engineering community discussion," which indicating a failure in understanding the context or intent of the question.

\begin{table}[htbp]
\caption{Error Categories with Their Frequency (Temperature 0.2, Prompt=1)}
\vspace{-0.2cm}
\centering
\begin{tabular}{|p{6cm}|c|}
\hline
\textbf{Category} & \textbf{Count} \\
\hline
Labeling error & 46\\
Absence of Explicit Offensive Language & 23 \\
Sarcasm and Irony & 16 \\
%Complex Linguistic Structures & 7 \\
Nuanced Toxicity & 6 \\
Non-Responsive Answer & 6\\
Context-Dependent Toxicity & 5 \\
Lengthy Phrasing & 4\\
\hline
\end{tabular}
\end{table}

\textbf{Labeling error:} We observed 46 instances of human error in labelling the dataset. These errors could be attributed to various factors, including the subjective nature of human annotation and possible misinterpretation of contextual cues. Additionally, some comments can be interpreted as a retaliation to a previous toxic comment (e.g., \textit{``Alright.... Nobody accused you 'falsely' you clearly were not respectful}"), or more subtle forms of behavior such as entitlement (e.g., ``Maybe when I state that the values in the extractor are correct, people should listen!!!"). Both  examples were labeled non-toxic by the annotators, but detected as toxic by ChatGPT. %the sheer volume of data to assess and the possibility of moderator fatigue or cognitive biases during the classification process.

%% file: conclusion.tex
\section{Discussion and Implications}
%Implications for OSS communities to curb toxicity
To determine the viability of ChatGPT for use by OSS communities in combating toxicity, we explored two key factors: (a) model interpretability, essential to provide meaningful insights and actionable information; and (b) potential avenues for integrating ChatGPT effectively into the software engineering workflow.

\subsection{Model Interpretability}
Model interpretability revolves around our ability to understand and explain the rationale behind the decisions made by ChatGPT. In the context of combating toxicity within OSS communities, interpretability ensures that the insights and information provided by the model are not only accurate but also comprehensible. This clarity empowers OSS practitioners to take informed actions and address toxicity effectively. 
We used the same dataset to explore a few prompts that could help us understand the classification rationale.  %We observed that asking for "a short justification" often is sufficient to gain some insights into the model outputs. 
Table \ref{prompts} shows sample outputs of such  justifications.

\begin{table}[htbp]
    \centering
     \caption{Sample Prompt Outputs for Toxicity Justification}
     \vspace{-0.2cm}
    \begin{tabular}{|p{8cm}|}
        \hline  	
        \rowcolor{aliceblue}\textbf{Prompt:} Examine whether the software engineering community discussion includes toxicity. Please provide your answer in yes/no with a short justification. \\ \hline \hline
            
            \textbf{Statement} - \textit{you're useless, please leave}\\
            \textbf{ChatGPT Output:}
            Yes. It includes personal attacks and a disrespectful tone. 
        \\\hline
        %\textbf{Prompt:} Examine whether the software engineering community discussion includes toxicity. Please provide your answer in yes/no with a short justification. 
            \textbf{Statement} - \textit{Pardon my English @friend, what you wrote is bullshit.}\\

            \textbf{ChatGPT Output:} 
            Yes. The use of the word "bullshit" is offensive, which can contribute to a toxic environment.
        \\\hline
        %\textbf{Prompt:} Examine whether the software engineering community discussion includes toxicity. Please provide your answer in yes/no with a short justification. 
        \textbf{Statement} - \textit{I'll report you too @friend, for all the internet to see how disgusting and abusive you are}\\

            \textbf{ChatGPT Output:} 
            Yes. The statement provided includes a threat to publicly shame and report someone, which can be seen as a form of toxicity. This kind of behavior is not constructive and can create a hostile environment within the community.'
        \\\hline
    \end{tabular}
    \label{prompts} 
\end{table}

\subsection{Integration into the Workflow} 
\begin{figure}
    \centering
    \includegraphics[width=0.48\textwidth]{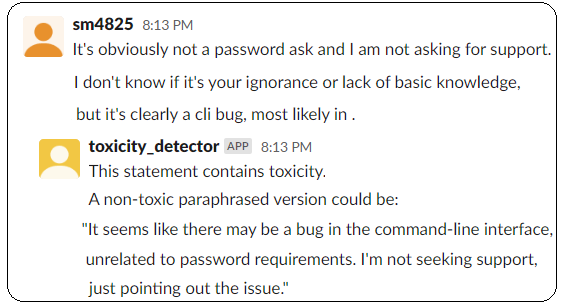} 
    \vspace{-0.6cm}
    \caption{Bot Implementation for Paraphrasing Toxic Text}
    \label{fig:image_label} 
\end{figure}

We explored the potential of developer chats as platforms for integrating preemptive interventions, as they facilitate spontaneous expression of emotions often not captured in other communication types~\cite{Chatterjee21, Kuutila2020ChatAI, DARE_23}. Specfically, we integrated ChatGPT into Slack, a popular chat platform, that would be able to automatically detect toxic content, and when necessary, generate an alternative, non-toxic version. This interface will detect harmful text, provide users with real-time feedback on the tone of their message, and suggestions to reframe their messages to convey positive and constructive intentions, as shown in Figure \ref{fig:image_label}. Additional details on the bot implementation is included in our replication package. 
%Additionally, when toxicity is detected in the input content, we employ prompts to guide OpenAI's GPT-3.5 Turbo in generating non-toxic paraphrased versions.The prompt structure for toxicity examination is as follows:

\vspace{-0.2cm}

%\subsection{Fine-Tuning for Software Engineering Communities.}  Explore the possibility of fine-tuning language models like GPT-3.5 Turbo on data specific to software developer communities~\footnote{\url{https://platform.openai.com/docs/guides/fine-tuning}}. Tailoring the model to the language and context commonly used within these communities could enhance the accuracy of toxicity detection.

\section{Conclusion and Future Work}
In this paper, we presented an approach for automated toxicity detection in software developer communication using a zero-shot LLM, namely ChatGPT, through a prompting approach. We evaluated GPT-3.5 Turbo on a previously curated dataset of GitHub issue comments, and observed promising preliminary results. We explored several prompts for toxicity detection, including those that outputs justification of model outputs. This holds particular significance in building trust among software engineers, encouraging their adoption and daily use of such systems in their workflows. We publish the source code and annotated dataset to facilitate the replication of our study at:  \url{https://anonymous.4open.science/r/open-source-toxicity-0236/README.md}.

There are several avenues for future work. First, our study focused only on one type of developer communication, i.e., issue comments on GitHub. Future studies should evaluate our approach on larger and more diverse datasets, such as code reviews or emails. Second, further work on prompt engineering could potentially help improve ChatGPT's performance on toxicity detection in software-related text. Third, there is a  possibility of fine-tuning language models like GPT-3.5 Turbo on data specific to software developer communities~\footnote{\url{https://platform.openai.com/docs/guides/fine-tuning}}. Tailoring the model to the language and context commonly used within these communities could potentially enhance the accuracy of toxicity detection. Overall, our study provides a starting point for future research to explore the potential of ChatGPT in detecting toxicity or incivil language in software engineering communication.